\documentclass{appolb}
\usepackage{graphicx}

\begin{document}
\title{Long-Lived Particles at Future Colliders.%
\thanks{Presented at XXVII Cracow EPIPHANY Conference on Future of particle physics}%
}
\author{Rebeca Gonzalez Suarez	
\address{Uppsala University, Sweden}
}
\maketitle
\begin{abstract}
Long-lived particles have significant enough lifetimes as to, when produced in collisions, leave a distinct signature in the detectors. Driven by increasingly higher energies, trigger and reconstruction algorithms at particle colliders are optimized for increasingly heavier particles, which in turn, tend to be short-lived. This makes searches for long-lived particles difficult, usually requiring dedicated methods and sometimes dedicated hardware top spot them. However, taking upon the challenge brings enormous potential, since new, long-lived particles feature in a variety of promising new physics models that could answer most of the open questions of the standard model, such as: neutrino masses, Dark Matter, or the matter-antimatter unbalance in the Universe. 

Currently, the international high energy physics community is planning future facilities post-LHC, and various particle colliders have been proposed. Crucial physics cases connected to long-lived particles will be accessible then, and in this presentation, three interesting examples are highlighted: Heavy Neutral Leptons, Hidden Sectors connected to Dark Matter, and exotic Higgs boson decays. This is followed by a small review of the preliminary studies assuming different future colliders, exploiting the complementary advantages that different colliding particles and accelerator types provide. 
\end{abstract}
\PACS{13.66.-a, 14.80.B, 14.60.St, 14.60.Pq, 29.17.+, 29.20.Lq }
  
\section{Introduction}
In the Standard Model (SM) of particle physics elementary and subatomic particles display different lifetimes. How long the lifetime of a particle is depends on different factors, mass and couplings being the most important. A particle with large mass, like the Z boson, will decay fast and a particle that only couples very feebly with others will have long decay times. 

The concept of ``long-lived particle'' in an experimental context accounts for the experimental setup itself. For example, whereas the Higgs boson, that has a lifetime of $10^{-22}$~s is certainly short-lived at the LHC, the muon, that takes just microseconds to decay, is in fact detector-stable since it has time to cross the whole volume of the LHC experiments during that time, and to all purposes, long-lived. 

It is then useful to understand ``long-lived particles'' as an umbrella term covering particles with lifetimes long enough to travel measurable distances inside the detectors before decaying, long enough to have distinct experimental signatures. And though both the above-mentioned muon and for example a photon, which is stable as far as we know, are then technically long-lived particles, the term is more commonly used to refer to new, beyond the Standard Model (BSM) particles that have not yet been observed experimentally. 

New, Long-Lived Particles (LLP) are in fact not a prediction of a single theory, and fit into virtually all proposed frameworks for BSM physics. Their presence is strongly motivated, and they typically are feebly interacting. Understanding the Nature of Dark Matter is among the different SM questions that LLP could give an answer to, together with Neutrino masses, Baryogenesis, and Naturalness to name a few; and they feature in different Supersymmetry (SUSY), compositeness, and hidden sector models. 

Searches for LLP are not new but they are gaining traction~\cite{Alimena:2019zri}, as an alternative and complement to more mainstream new physics searches, and will be at least as interesting, if no more, at future colliders. Moreover, models featuring LLP can potentially be tested not only at different colliders, but by astro-particle and non-collider (mainly accelerator) experiments, making them ideal for cross-disciplinary, collaborative work. 

\section{Three interconnected physics cases}

In this section, three possible physics cases of LLPs  that could be fully explored at future colliders are presented. 

\subsection{Heavy Neutral Leptons}

The discovery of Heavy Neutral Leptons (HNL) could answer central open questions of the SM, conceivable more than one at the same time. Just to pick one example, in the $\nu$MSM~\cite{Asaka:2005pn, Shaposhnikov:2006xi}, light sterile neutrinos can accomplish leptogenesis, a mechanism to dynamically generate the matter-antimatter asymmetry of the Universe via HNL decays during the first $10^{-27}$~s after the Big Bang; provide a Dark Matter  candidate; and solving the question on how neutrino acquire mass. 

\begin{figure}[htb]
\centerline{%
\includegraphics[width=12.5cm]{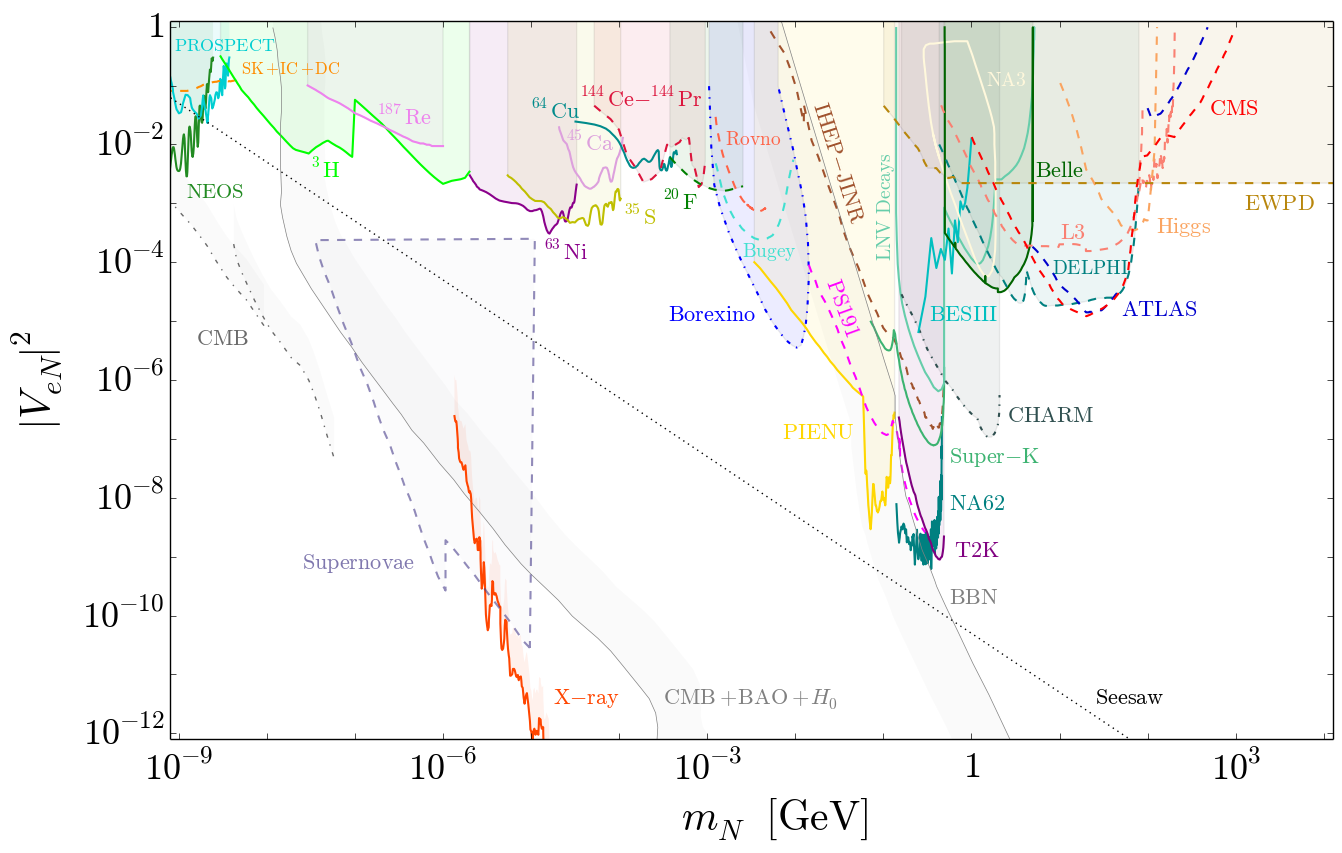}}
\caption{Current constraints on the electron neutrino-sterile neutrino mixing $|V_{eN}|^2$ as a function of the sterile neutrino mass $m_{N}$, from Ref.~\cite{Bolton:2019pcu}.}
\label{Fig:F1}
\end{figure}

It is then easy to understand why HNL are among the most interesting new physics to target not only at particle colliders, but also at astrophysics and non-collider experiments. Most of the current limits, as presented for example in Fig.~\ref{Fig:F1}, cover high neutrino mixing values. For low values of the neutrino mixing angle, most of the still available phase space, the decay length of the heavy neutrino can be significant. This means that HNL with low mixing angles are in fact LLP. For example, HNL produced in Z boson decays as $\mathrm{Z}\rightarrow\nu\mathrm{N}$ decaying into $\mathrm{N}\rightarrow l\mathrm{W}$ will produce a displaced vertex with multiple tracks. For HNL with long enough lifetimes, oscillations can also be studied~\cite{Antusch:2017ebe}.

\subsection{Hidden Sectors}

At the moment, we know there is more to Nature than the SM, but as stated in the introduction, we have no way of predicting the energy scale where new physics may exist. The concept behind Hidden Sectors is that full collections of new particles and forces could be already accesible by the current experimental energies, but just not interacting or doing it very feebly with the SM particles. If very small couplings, called ``portals'', of these new collections of particles and forces with SM particles existed, the Higgs boson being one of the prime candidates, that would produce subtle experimental signs to follow. Small couplings lead to long-lived signatures.

Within the Matter-Energy content of the Universe, Dark Matter represents about 27\%, while regular matter made up of SM particles just about 5\%. In other words, Dark Matter makes up $>$80\% of all matter. Regular matter is very complex and it would not be far-fetched to assume a similar complexity for Dark Matter. Thus, Hidden Sectors connected to Dark Matter, also called ``Dark Sectors'', are an attractive area of research in the current experimental landscape of Dark Matter searches in direct, indirect, and collider experiments. In the context of Dark Sectors there are many long-lived signatures that arise, some of them quite exciting, like, e.g.: Dark Showers. 

``Axion-Like Particles'' (ALPs) provide a very-weakly-coupled window to the Dark Sector that can be explored widely at future colliders. For small couplings and light ALPs, the ALP decay vertex can be considerably displaced from the production vertex, that is, conforming a long-lived signature~\cite{Bauer:2018uxu}. 
 
 \begin{figure}[htb]
\centerline{%
\includegraphics[width=12.5cm]{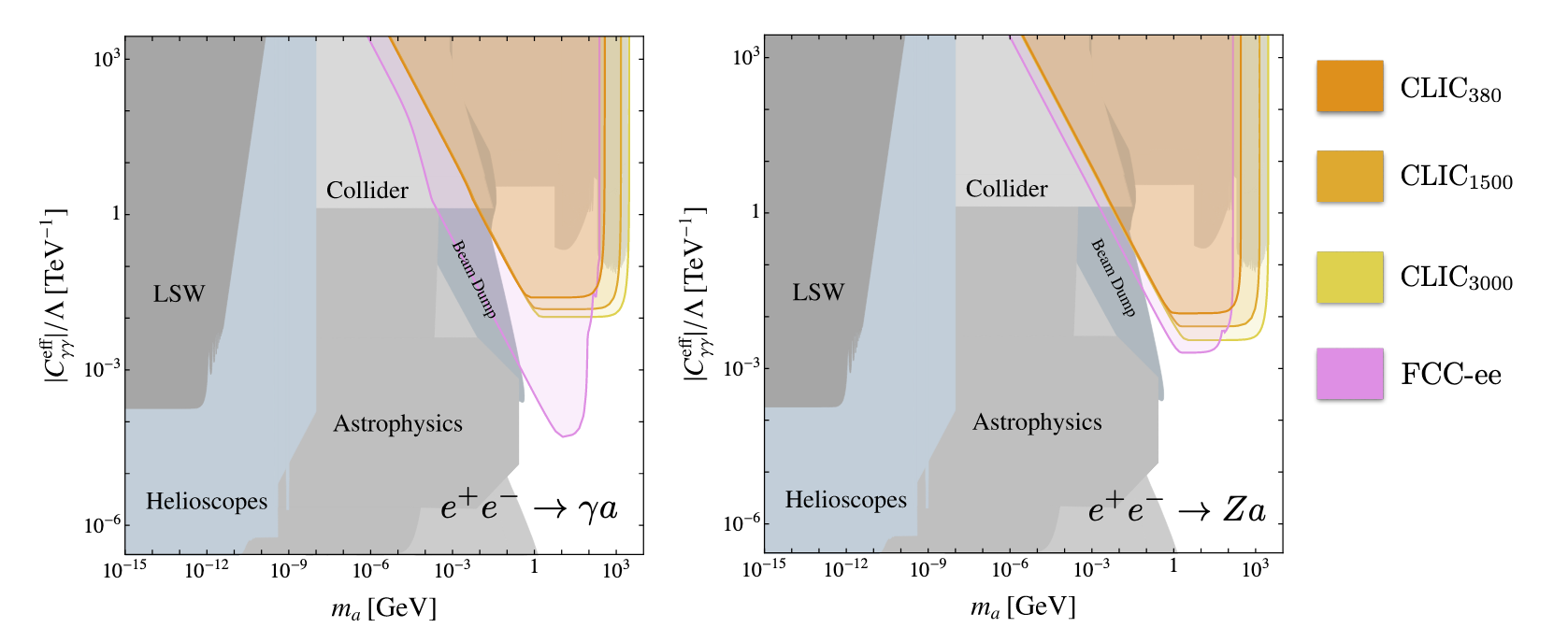}}
\caption{Projected sensitivity regions for searches for $e^+ e^- \rightarrow\gamma a \rightarrow 3\gamma$ (left) and $e^+ e^- \rightarrow Z a \rightarrow Z_{vis}\gamma\gamma$ (right) at future $e^+e^-$ colliders for BR($a\rightarrow\gamma\gamma$) = 1.  The constraints from Figure 4 are shown in the background. The sensitivity regions are based on 4 expected signal events. From Ref.~\cite{Bauer:2018uxu}.}
\label{Fig:F2}
\end{figure}
 
\subsection{Exotic Higgs boson decays}\label{exohiggs}

 We are still getting to know the Higgs boson. At the moment, measurement and searches characterize a Higgs boson largely compatible with he SM, but the uncertainties are large in some measurements and there could still be exotic behaviour to be found, that would provide indications of what can lie beyond the SM. This is one of the reasons why a Higgs factory is the first priority for the European Strategy for particle physics~\cite{CERN-ESU-015}. 
 
 Within the exploitation of the Higgs sector, exotic Higgs boson decays to LLPs can be explored at future lepton colliders. This kind of decay features in different BSM models, some related to Hidden sectors, like Twin Higgs or
Hidden Valley models~\cite{Alipour-Fard:2018lsf}. 

\section{Technical challenges}

High-energy particle colliders push the Energy Frontier, producing heavier particles with each increase of the center of mass energy. Heavier particles tend, in exchange, to be shorter-lived.  A primer example of this is the Higgs boson at the LHC. A Higgs boson is produced and decays about $10^{-22}$ seconds after. It often decays in turn to heavy particles, for example, to two Z bosons, each of them decaying in about $10^{-25}$~s in e.g. a couple of muons each, giving way to one of the clearest channels for discovery. In practice this means that 4 muons coming from the collision point are observed. The LHC detectors, trigger, and reconstruction methods are designed to find this kind of signature. 

New, LLPs are not like that, and display a collection of different signatures which are quite unique. From the most intuitive displaced tracks and vertices to unconventional emerging signatures passing by disappearing or kinked tracks; tracks with anomalous energy distributions; or slow or stopped particles that appear out of time. 

The implications of this are various. First of all, it means that, by displaying uncommon signatures, LLP are affected by little to no background. The downsides of this are: potential instrumental background that could be difficult to model, and the fact that they may require dedicated reconstruction techniques, trigger algorithms, and even dedicated detector design to be identified. 

This is a problem at current high and low energy colliders, and will not be different at future colliders, independently of accelerator or particle types. At this point, when discussing future facilities, one could design the future detectors as usual and then try to make the best out of them for LLPs, which presents difficulties, as we know from the current collider experience; or LLPs could be included in the design strategy for future collider detectors, prioritizing for example displaced tracking capabilities, or timing, and budgeting for unexpected signals. Following the second path will result in a boost of the discovery potential for LLP searches; but could also have additional benefits providing innovative methods. 

\begin{figure}[htb]
\centerline{%
\includegraphics[width=9cm]{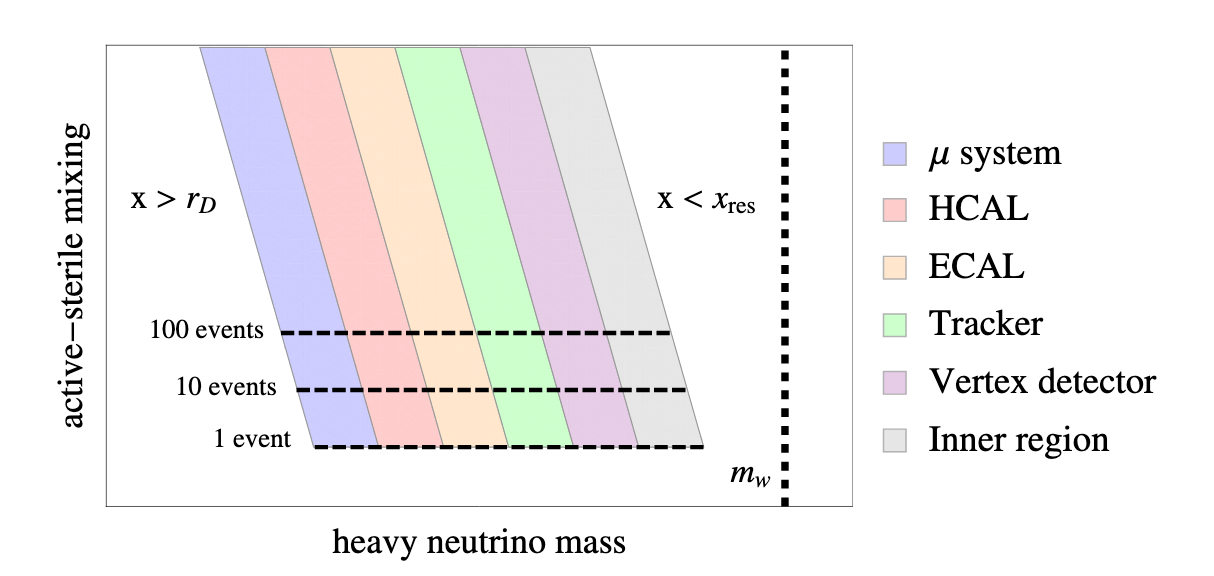}}
\caption{Schematic illustration of the sensitivity of the different detector components to heavy neutrino decays as a function of the active-sterile mixing parameter and the heavy neutrino mass. The parameter $r_D$ the outer radius of the muon system. From Ref.~\cite{Antusch:2016vyf}.}
\label{Fig:F3}
\end{figure}

Figures like Fig.~\ref{Fig:F3}, that presents the sensitivity of different detector components to HNL as a function of the mixing parameter and mass, from Ref.~\cite{Antusch:2016vyf}, could provide important input for the detector design of the future. 

Different colliders offer sensitivity to different parameters in different models, which means that complementarity should be exploited. Certain proposed colliders offer high collision energies, other clean environments, with different access to ranges of mass and couplings. There is also complementarity with non-collider and astrophysics in many models. However, many challenges related to LLP will be common. There is then a clear benefit in developing searches together and establish common benchmarking. 

\section{Future particle colliders}

\subsection{The High Luminosity Upgrade of the LHC}\label{hilumi}
 \begin{figure}[htb]
\centerline{%
\includegraphics[width=12.5cm]{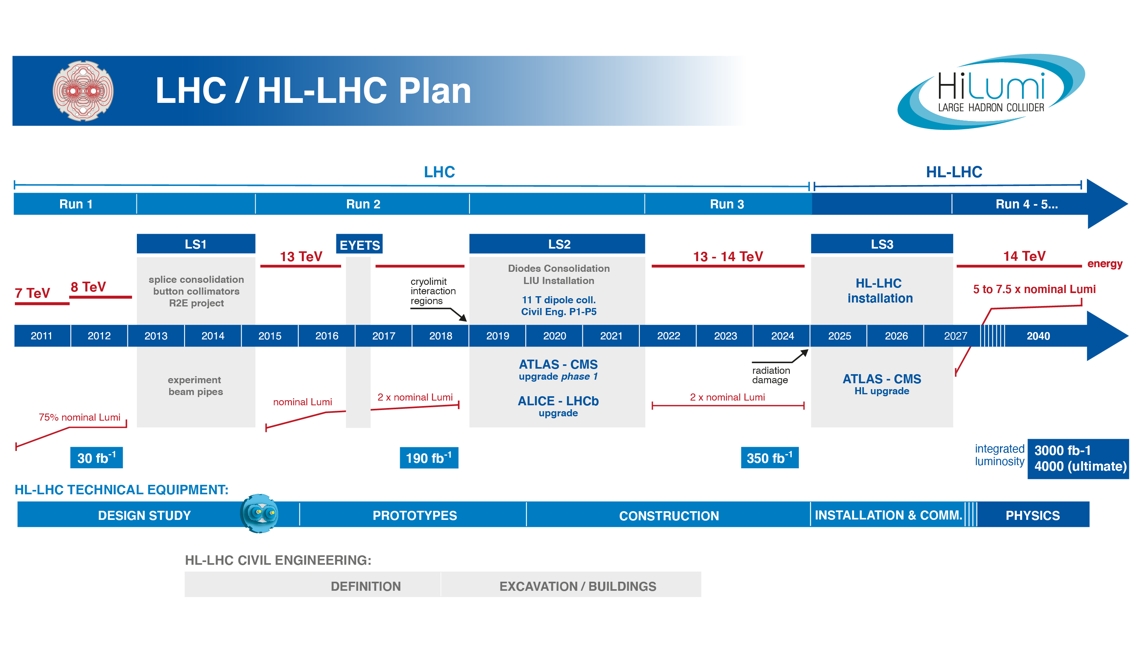}}
\caption{LHC/ HL-LHC Plan (last update August 2020). From~\cite{hllhc}}
\label{Fig:F4}
\end{figure}

The LHC is currently undergoing its second, planned, long shut-down (LS2). In order to further increase its discovery potential after the next running period of the LHC, the current accelerator chain would eventually undergo an upgrade. The High Luminosity LHC (HL-LHC) is an upgrade of the LHC to achieve instantaneous luminosities a factor of five larger than the LHC nominal value, thereby enabling the experiments to enlarge their data sample by one order of magnitude compared with the LHC baseline programme. Following five years of design study and R\&D, this challenging project requires now about ten years of developments, prototyping, testing and implementation; hence operation is expected to start in the middle of the next decade. 

The timeline of the project is dictated by the fact that, at the beginning of the next decade, many critical components of the accelerator will reach the end of their lifetime due to radiation damage and will thus need to be replaced. The upgrade phase is therefore crucial not only for the full exploitation of the LHC physics potential, but also to enable operation of the collider beyond 2025~\cite{doi:10.1142/9581}. The current schedule of the project is presented in Fig.~\ref{Fig:F4}.

New functionalities of the HL-LHC could have positive effects in searches for LLPs. For example, the new track triggers~\cite{Martensson:2019sfa} that the LHC experiments will be equipped with could allow for example to trigger on displaced muons from the same vertex to find dark photons~\cite{Gershtein:2017tsv}. Better timing information will be available, and that would enable e.g. searches that target pair-produced LLPs significantly delayed~\cite{Liu:2018wte}. 

A very exciting development related to the HL-LHC is a variety of dedicated LLP detectors proposed around it. This complementary instrumentation in the caverns would offer low background environments to detect LLPs. 
\begin{itemize}
\item FASER~\cite{Feng:2017uoz}, already approved, is a 1~$m^{3}$ on-axis experiment, located 480~m downstream from the interaction point of the ATLAS experiment. 
\item MATHUSLA~\cite{Curtin:2018mvb}, is a proposed large-scale surface detector instrumenting about~$8\times10^5$~$m^3$, off-axis, above ATLAS or CMS.
\item CODEX-b~\cite{Aielli:2019ivi}, is a proposed $10^3$~$m^3$ detector in the LHCb cavern, off-axis.
\item AL3X~\cite{Gligorov:2018vkc}, is a proposed cylindrical  $900$~$m^3$ detector inside the L3 magnet and the time-projection chamber of the ALICE experiment.
\item ANUBIS~\cite{Bauer:2019vqk}, proposes $1\times1$~$m^3$ units on top of ATLAS or CMS, off-axis.
\end{itemize}

At the moment of writing, the HL-LHC is the only approved future collider. However, there are many proposed future options beyond that and there will be briefly presented, together with some selected preliminary search proposals for LLPs, in the next subsections. 

\subsection{Lepton Colliders}

As stated in section~\ref{exohiggs}, the European Strategy Update highlights the need to pursue an electron positron collider acting as a Higgs factory as the highest-priority facility after the LHC. For LLPs this means the clear physics case discussed previously of exotic Higgs boson decays~\cite{Cheung:2019qdr}, but that is only the starting point. 

There are two kinds of proposed electron-position colliders: Linear and Circular. In terms of linear electron-position colliders, two projects can be highlighted: CLIC~\cite{Robson:2018enq} at CERN, and the ILC~\cite{Bambade:2019fyw} in Japan. Hidden valley searches in Higgs boson decays with displaced vertices~\cite{Kucharczyk:2625054} or Degenerate Higgsino Dark Matter via chargino pair production searches with disappearing tracks~\cite{deBlas:2018mhx} are proposed to be performed at CLIC.

In terms of circular electron-positron colliders, the options are the CEPC~\cite{CEPCStudyGroup:2018rmc} in China, and the FCC-ee~\cite{Abada:2019zxq} at CERN, the lepton collider proposed within the Future Circular Collider. 

The FCC-ee offers exciting potential for the study of LLP, where searches can be not only complementary to similar searches at collider and non-collider experiments, but highly competitive. More specifically, the FCC-ee offers an unbeatable reach for HNL at the Z-Pole, making it the flagship of LLP searches in this collider. 

Circular electron-positron colliders are expected to provide the best sensitivity to low neutrino mixing angles via displaced  vertex  searches~\cite{Antusch:2016ejd}. The large statistics FCC run around the Z pole, producing 5 $10^{12}$ Z (the Tera-Z regime), is expected to be particularly powerful in this area~\cite{Blondel:2014bra,Abada:2019zxq,Klaric:2020lov,Abada:2018oly}. The Tera-Z run would allow for sensitivity down to a heavy-light mixing of $10^{-11}$, covering a large phase-space for heavy neutrino masses between 5 (the B mass) and 80~GeV (the Z mass) with displaced vertex searches. It is important to note that this kind of searches are affected by very little background, since the displacement of the vertex can be e.g. of 1~m. Sufficiently long-lived HNL could also potentially allow for oscillations into antineutrinos to be observed~\cite{Antusch:2017ebe}.

\begin{figure}[htb]
\centerline{%
\includegraphics[width=7.5cm]{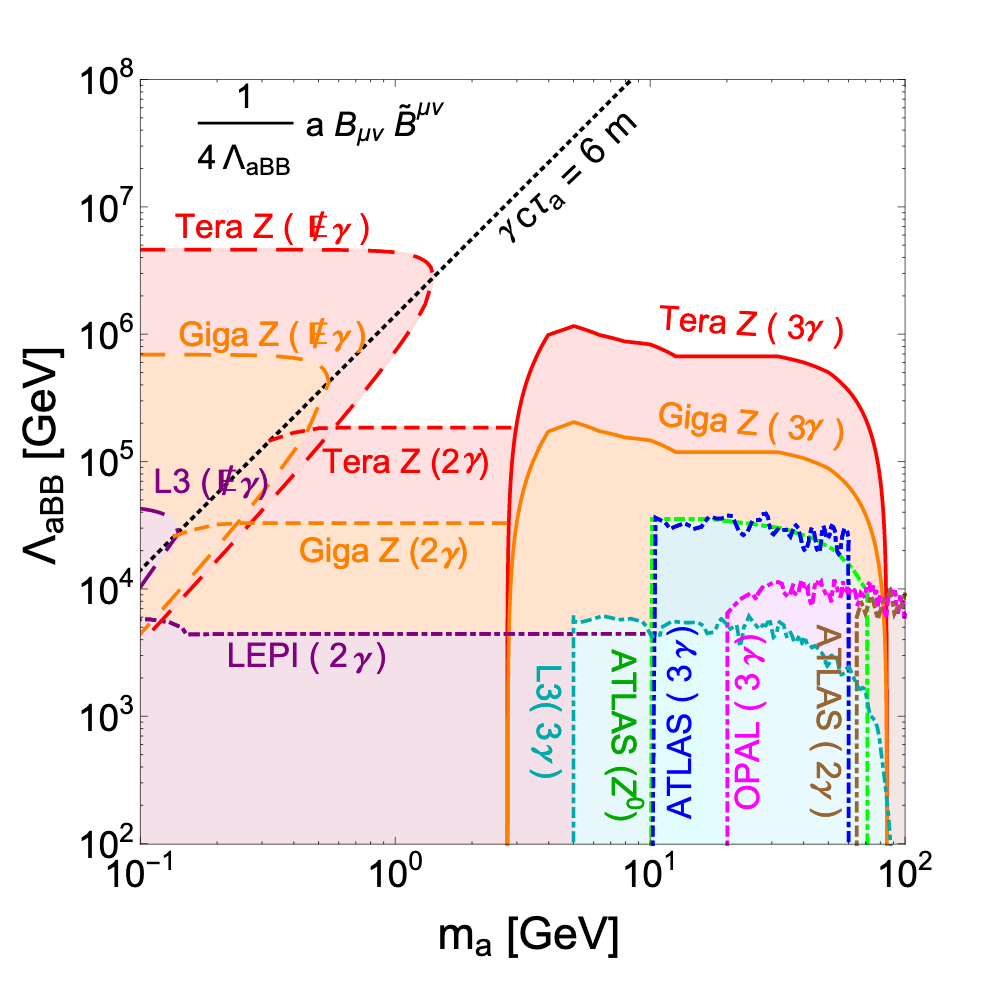}}
\caption{The limit on $\Lambda_{aBB}$, ALP coupling to hypercharge field, from future Z-factory. More details at the source, Ref.~\cite{Liu:2017zdh}.}
\label{Fig:F5}
\end{figure}

Evidence for a Dark Sector at the FCC-ee could come in the form of ALPS. For small couplings and light ALPs, the ALP decay vertex can be considerably displaced from the production vertex. Very long-lived ALPs would leave the detector before decaying, leaving a trace of missing energy. As presented in Fig~\ref{Fig:F5}, and it has been shown in~\cite{Bauer:2018uxu}, a high-luminosity run at the Z pole would significantly increase the sensitivity to ALPs produced in $e^+e^-\rightarrow \gamma a$ with subsequent decays $a\rightarrow \gamma\gamma$ or $a \rightarrow l^+l^-$.  

\begin{figure}[htb]
\centerline{%
\includegraphics[width=7.5cm]{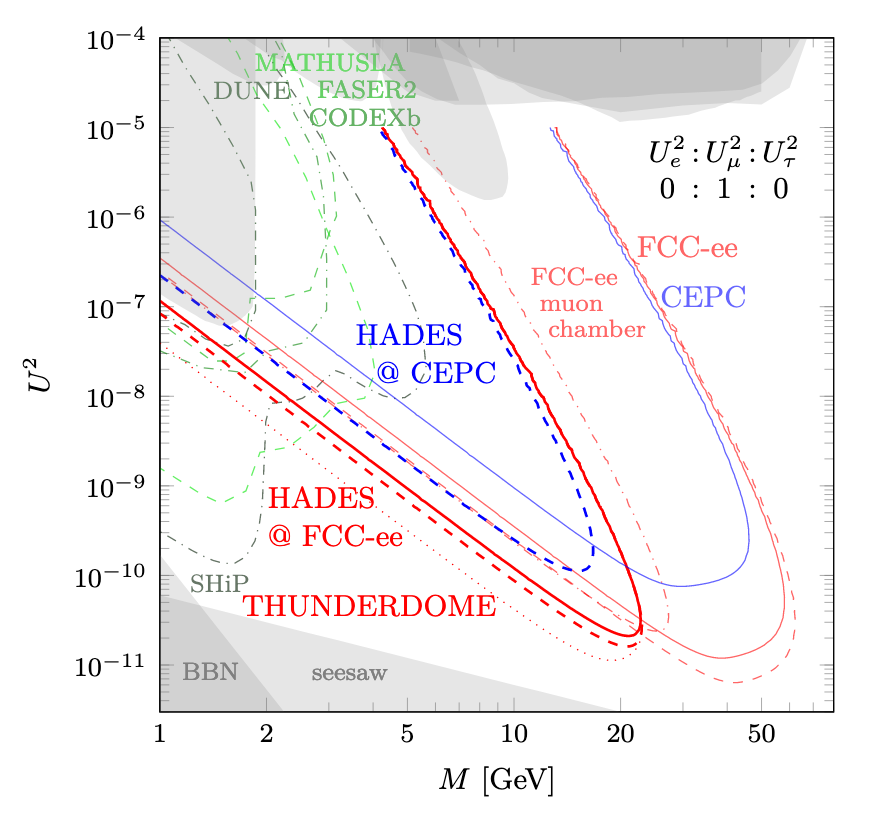}}
\caption{Comparison of the sensitivities (9 events) that can be achieved at the FCC-ee with $2.5\times10^{12}$ Z-bosons (red) or CEPC with $3.5\times10^{11}$ Z-bosons (blue). More details at the source, Ref.~\cite{Chrzaszcz:2020emg}.}
\label{Fig:F6}
\end{figure}

Following the plans for different additional LLP experiments at the HL-LHC presented in~\ref{hilumi} it is possible to also envision similar concepts at other future colliders. This is the premise for HADES: A long lived particle detector concept for the FCC-ee or CEPC~\cite{Chrzaszcz:2020emg}. The civil engineering of the FCC-ee will have much bigger detector caverns than needed for a lepton collider in order to use them further for a future hadron collider. Extra instrumentation could then be installed in the cavern walls to search for new LLPs. Figure~\ref{Fig:F6} shows the improvement of sensitivity that can be expected in HNL searches at the FCC-ee  or CEPC by HADES.

Finding valid motivations for LLP at the FCC-ee/CEPC is certainly not a problem. As previously discussed it would be possible to explore Twin Higgs models with displaced exotic Higgs boson decays, Hidden Valley models with neutral, long-lived particles that the Higgs boson can decay to~\cite{Alipour-Fard:2018lsf}. Other BSM searches that could be competitive concern dark glueballs~\cite{Cheung:2019qdr}; Neutral naturalness~\cite{Curtin:2015fna}; or Neutralinos~\cite{Wang:2019orr} among others. 

\subsection{Future Hadron colliders}

Several hadron colliders and electron-hadron colliders have been proposed: the LHeC at CERN~\cite{Kuze:2018dqd}; the HE-LHC also at CERN~\cite{Todesco:2011np}; the SppC in China~\cite{Tang:2015qga}; and the FCC-eh/hh at CERN~\cite{Kuze:2018dqd,Benedikt:2018csr}.

\begin{figure}[htb]
\centerline{%
\includegraphics[width=7.5cm]{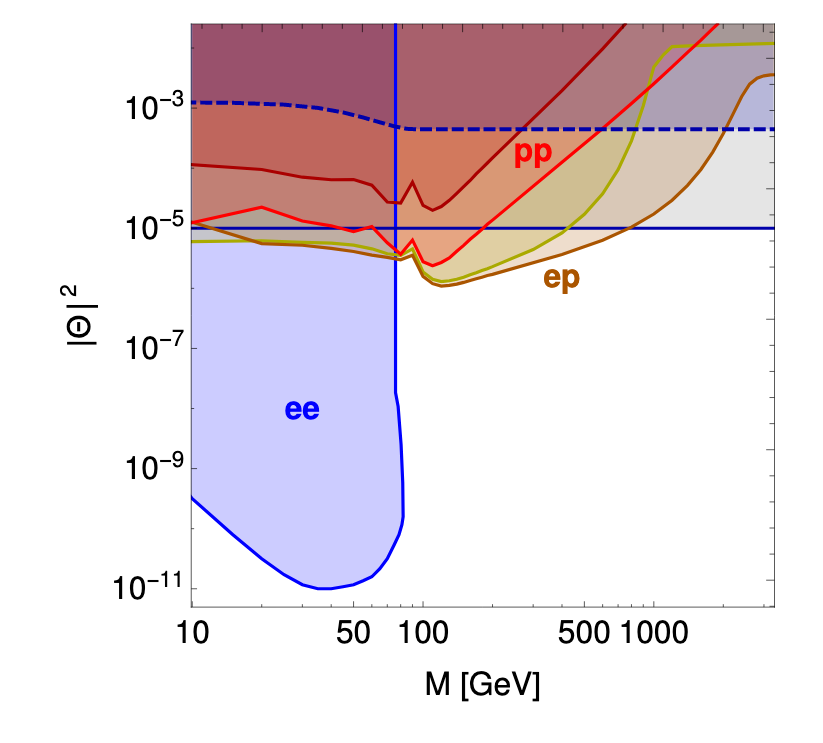}}
\caption{Summary  of  selected  estimated  sensitivities  of  the  FCC-ee, -hh, and -eh colliders, including the HL-LHC and the LHeC. The best sensitivity for heavy neutrino masses $M < m_{W}$ is obtained from the displaced vertex searches at the Z pole run of the FCC-ee shown by the blue line. More details at the source, Ref.~\cite{Antusch:2016ejd}.}
\label{Fig:F7}
\end{figure}

Among the available studies of LLPs in future hadron and lepton-hadron colliders, there are a few examples to highlight. One of them is the potential for searches for Dark Photons in electron-hadron colliders~\cite{DOnofrio:2019dcp} that could be quite competitive. LLP signals arising from Higgsinos or exotic Higgs decays can also have good potential at lepton-proton colliders~\cite{Curtin:2017bxr}.  Wino and higgsino dark matter searches  with a disappearing track signatures have been proposed for the FCC-hh~\cite{Saito:2019rtg}. Finally HNL, Sterile Neutrinos in particular, can be explored  at future hadron colliders as well~\cite{Antusch:2016ejd}, and the complementarity between $e^-e^+$, $pp$, and $e^-p$ in these flagship searches is an asset to take in consideration, as presented in Fig.~\ref{Fig:F7}. 

\section{Conclusions and summary}

Searches for long-lived particles are a very attractive complement to mainstream new physics searches at colliders. However, this kind of searches also challenge conventional reconstruction and trigger methods. 

There are many interesting lines involving new, long-lived particles to explore at future colliders, some of them very well-known, like Heavy Neutral Leptons, Hidden sectors that could be connected to Dark Matter, or exotic Higgs decays. While in general most proposed future colliders can access many of these models, with common challenges, different colliders access different areas of the phase space and offer different sensitivities. In order to answer the big questions of the SM, synergies across experimental facilities should be maximized beyond colliders. 
 
Now is the time to organize and properly benchmark the most interesting options and  dive-in into dedicated detector design; not only to just being able to reconstruct and identify long-lived particles at future colliders, but to maximize the experimental coverage and fully exploit the available discovery opportunities.

\bibliographystyle{ieeetr}
\bibliography{references}

\end{document}